\documentclass[notitlepage, amsmath, nofootinbib, aps, reprint,groupedaddress]{revtex4-1}
\usepackage[utf8]{inputenc}
\usepackage{txfonts}
\usepackage{xcolor}
\usepackage{graphicx}
\usepackage{bm}
\usepackage{ulem} 

\begin{document}
\newcommand{\dgw}{{d_{\rm GW}}}
\newcommand{\dgwtrue}{{d_{\rm GW,true}}}
\newcommand{\dgwobs}{{d_{\rm GW,obs}}}
\newcommand{\dem}{d_{\rm EM}}
\newcommand{\dl}{d_{\rm L}}
\newcommand{\zmin}{{z_{\rm min}}}
\newcommand{\zmax}{{z_{\rm max}}}
\newcommand{\kmsMpc}{\,{\rm km/s/Mpc}}
\newcommand{\Mpc}{\,{\rm Mpc}}
\newcommand{\ztrue}{{z_{\rm true}}}
\def\hinvmpc{\,h^{-1}{\rm Mpc}}
\newcommand{\ds}{\displaystyle}
\def\hinvMsun{\,h^{-1}{\rm \Msun}}

\newcommand{\Msun}{\, \rm M_\odot}
\newcommand{\hiMsun}{\,h^{-1}\rm M_\odot} 
\newcommand{\Mvir}{M_{\rm vir}}
\newcommand{\hikpc}{\,h^{-1}\rm kpc}
\newcommand{\hiMpc}{\,h^{-1}\rm Mpc}
\newcommand{\hiGpc}{\,h^{-1}\rm Gpc}

\newcommand{\emery}[1]{\textcolor{red}{ET: #1}}
\newcommand{\dragan}[1]{\textcolor{violet}{DH: #1}}
\title{Challenges for the statistical gravitational-wave method to measure the Hubble constant
}

\author{Emery Trott} \email{etrott@umich.edu} 

\author{Dragan Huterer} \email{huterer@umich.edu} 

\affiliation{Department of Physics, University of Michigan, 450 Church St, Ann Arbor, MI 48109-1040}
\affiliation{Leinweber Center for Theoretical Physics, University of Michigan, 
450 Church St, Ann Arbor, MI 48109-1040}

\begin{abstract}
Gravitational waves (GW) can be employed as standard sirens that will soon measure the Hubble constant with sufficient precision to weigh in on the $\sim 5\sigma$ Hubble tension. Most GW sources will have no identified electromagnetic counterpart, leading to uncertainty in the redshift of the source, and in turn a degeneracy between host galaxy distance, redshift, and $H_0$. In the case where no electromagnetic counterparts are identified, it has been proposed that a statistical canvassing of candidate GW hosts, found in a large galaxy survey for example, can be used to accurately constrain the Hubble constant. We study and simulate this ``galaxy voting" method to compute $H_0$. We find that the Hubble constant posterior is in general biased relative to the true value even when making optimistic assumptions about the statistical properties of the sample. Using the MICECAT light-cone catalog, we find that the bias in the $H_0$ posteriors depends on the realization of the underlying galaxy sample and the precision of the GW source distance measurement.
\end{abstract}
\maketitle

\section{Introduction}

Gravitational-wave events are an emerging powerful probe that can be used to measure distances to inspiraling black holes or neutron stars. In the ``standard siren" technique \cite{Schutz:1986gp,Holz:2005df}, a combination of the distance measurement to the gravitational wave (GW) event and redshift to the measured electromagnetic (EM) counterpart can be used to determine the luminosity distance to the object. This can be further turned into a measurement of the matter and dark-energy densities, as well as that of the Hubble constant $H_0$. Such a measurement would be particularly useful, as current CMB measurements currently give $H_0 = (67.4\pm0.5)\kmsMpc$ \cite{Aghanim:2018eyx}, while low-redshift measurements based on the distance ladder give $H_0 = (73.04\pm1.04)\kmsMpc$ \cite{Riess:2021jrx}, implying the $H_0$ tension of 5.0$\sigma$. Standard-siren measurements have the potential to shed significant new light on the Hubble tension.

Currently, the main challenge with the standard-siren method is  the low number of merger events with a detected EM counterpart. LIGO has detected 50 mergers where at least one of the objects is a black hole, but only one of them, the binary neutron star system GW170817, has an EM counterpart that has been identified and measured \cite{LIGO:2017}.  The situation is expected to dramatically evolve in the next few years largely due to sensitivity upgrades to LIGO, and we can expect $\sim$100 binary neutron star detections by around 2025 \cite{Chen:2017rfc}. Some of these events will be close enough to be able to identify a single EM counterpart, in which case getting a measurement of $H_0$ is relatively straightforward.

However, the majority of distant GW events (which will outnumber nearby events due to volume considerations) will be poorly localized, leading to uncertainty as to which of the tens of thousands of potential host galaxies is the true host. In these cases we may employ a probabilistic method that allows the potential hosts to ``vote" on the preferred $H_0$, each host being weighted by its position relative to the center of the localization region \cite{Finn:1994cg,MacLeod:2007jd,Messenger:2011gi,DelPozzo:2012zz}. It has been argued that such an approach --- which we heretofore refer to as the \textit{statistical GW method} --- leads to an unbiased measurement of the Hubble constant \cite{MacLeod:2007jd,Nair:2018ign,Chen:2017rfc,Gray:2019ksv}. Such a method has already been applied in practice \cite{Soares-Santos:2019irc,Palmese:2020aof}. The statistical method's constraints on $H_0$ are currently weak, but are forecasted to become much tighter in the future \cite{Chen:2017rfc}.

In this paper we examine the statistical standard-siren method in some detail. We point out that, in the absence of clustering information in the source distribution, there are irreducible degeneracies between the parameters that enter the determination of the Hubble constant. We find that these degeneracies generally lead to biases in the derived value of $H_0$ that are only alleviated as conditions approach the bright siren limit.


\section{Statement of the problem}\label{sec:statement}

The waveform of a GW event enables a measurement of the luminosity distance, which is given by
\begin{equation}
    d_L(z,H_0) =  \frac{c(1+z)}{H_0} \int_0^z \frac{{\rm d}z'}{\sqrt{(1-\Omega_M)+\Omega_M(1+z')^3}}
    \label{eq:dL}
\end{equation}
where we have assumed a flat $\Lambda$CDM universe. Here $\Omega_M$ is the matter density relative to critical, $z$ is the redshift of the GW host galaxy, and $c$ is the speed of light. At $z\ll 1$, the equation simplifies to $d_L\simeq cz/H_0 + O(z^2)$, where $\Omega_M$ only comes in at the $z^2$ order in the term in parentheses; in this limit, the luminosity distance depends only on $z$ and $H_0$, and not on other cosmological parameters. In what follows, we retain the general expression in Eq.~(\ref{eq:dL}), and fix $\Omega_M=0.3$, but we occasionally come back to the simplified first-order expansion of $d_L(z)$ to make pedagogical points.

Equation (\ref{eq:dL}) showcases a fundamental challenge to measuring $H_0$, namely, for a given measurement of $d_L$ there is a perfect degeneracy between $H_0$ and $z$. That is, even with an infinite-precision $d_L$ measurement to a GW event, galaxies at higher redshift than the true host will be reporting a higher-than-true Hubble constant, and vice-versa for galaxies at lower $z$ than the actual host. In the statistical GW method, the host redshift is not known, and one rather marginalizes over redshifts of many thousands of potential  GW hosts, each of which is at a higher or lower redshift than the true host(s). 
Mathematically,  individual posteriors on $H_0$ from each of the many thousands of potential-host galaxies are averaged with the hope that they will average in just such a way to leave an unbiased measurement of $H_0$. In this paper, we study the statistical GW method, and show that it will be challenging to obtain such an unbiased measurement even under optimistic assumptions about some of the physical variables that enter the mathematical modeling of the method.

\section{Methodology}\label{sec:method}

To set up an analysis pipeline we need three ingredients: the electromagnetic (EM) data consisting of galaxies that could be potential GW hosts; the GW data consisting of galaxies that \textit{are} the hosts, and the statistical analysis procedure. We now describe them in that order.

\subsection{EM Data}\label{sec:em_data}

We adopt the public release of the  MICE-Grand Challenge Galaxy and Halo Light-cone Catalog (MICECAT), which is based on MICE-Grand Challenge (GC) simulations\footnote{http://maia.ice.cat/mice/} \citep{Fosalba:2013wxa,Crocce:2013vda,Fosalba:2013mra}.
The light-cone catalog was generated using a hybrid halo occupation distribution and halo abundance matching prescriptions to populate friends-of-friends dark matter halos from the MICE-GC simulation \cite{Carretero:2014ltj}. 
The input cosmological model to the MICE-GC simulation is a spatially flat model with matter density relative to critical $\Omega_M=0.25$ baryon density $\Omega_B=0.044$, amplitude of mass fluctuations $\sigma_8 = 0.8$, scalar spectral index $n_s=0.95$, and the scaled Hubble constant $h=0.7$. The catalog was built to observe local observational constraints on the luminosity function, galaxy clustering as a function of luminosity and color, and the color-magnitude diagram. It approximately reproduces the magnitude limits in the Dark Energy Survey. Host halos in MICECAT have masses $M > 2.2\times 10^{11} \hinvMsun$; we refer to them as ``galaxies" in what follows.

We access the MICECAT catalogs using \texttt{CosmoHub}\footnote{https://cosmohub.pic.es/home} web interface \citep{Tallada:2020qmg,Carretero:2017zkw}. We use the interface to select two sub-volumes on the MICECAT "sky", which we refer to as ``direction 1" and ``direction 2". We include galaxies coming within an opening angle $\theta$ of either $1^\circ$ or $5^\circ$ around the central direction of each of the corresponding volumes. We further downsample the catalog by selecting one out of 32 galaxies available in each volume. This leads respectively to $\sim$1,500 galaxies per volume (for $\theta = 1^\circ$), and $\sim$40,000 galaxies per volume (for $\theta = 5^\circ$). The two directions and two opening angles allow us to observe, respectively, the effects of sample variance and the size of the catalog on our final constraints on $H_0$.
In all cases the galaxy distribution extends to $z\simeq 1.4$, and is volume-limited out to $z\simeq 1$.

\begin{figure}[t]
    \includegraphics[width=\linewidth]{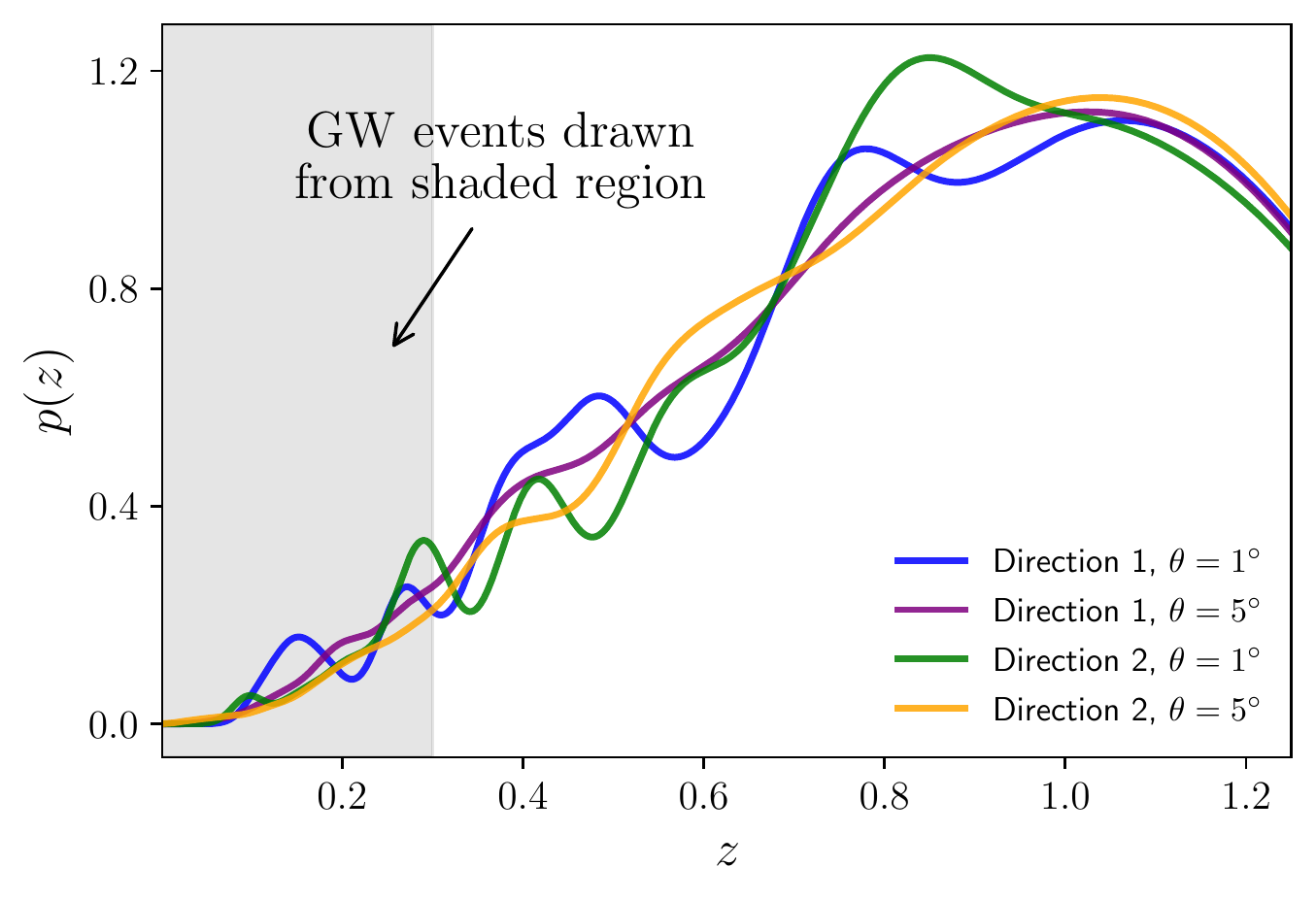}
    \caption{The photometric redshift distributions of four MICECAT light-cones. The volume-limited regime far exceeds the region from which GW events are drawn. The location of peaks in the shaded region of $p(z)$ end up corresponding to peaks in $p(H_0)$, provided the distance uncertainty is sufficiently low.}
    \label{fig:pz}
\end{figure}

Each galaxy in the sample has a maximum likelihood redshift value $z_i$. We assign each a redshift error $\sigma_{z_i} = 0.013(1+z_i)^3 \leq 0.015$, consistent with \cite{Soares-Santos:2019irc}, and compute its normalized photometric redshift distribution
\begin{equation}
    p_i(z) = \frac{1}{\sigma_{z_i}\sqrt{2\pi}} \exp \bigg{[}-\frac{1}{2}\bigg{(}\frac{z_i-z}{\sigma_{z_i}}\bigg{)}^2\bigg{]} \ .
    \label{eq:p_i}
\end{equation}
Then we sum to obtain our final voting distribution
\begin{equation}
    p(z) = \frac{1}{N_{\rm gal}} \sum_i^{N_{\rm gal}} p_i(z).
    \label{eq:pz}
\end{equation}
The quantity $p(z)$ is shown in Fig.~\ref{fig:pz} for the two directions and two opening angles. The fluctuations in these distributions are critical to determining the peak of the posterior. This is because the bumps and wiggles in $p(z)$ provide features in redshift that help break the degeneracy between redshift and $H_0$; see Sec.~\ref{sec:concl} for a more detailed discussion.

\subsection{GW Data}

We assign GW events randomly to galaxies from our EM sample, weighted by $p(z)$ and restricted to the range $0 < z < 0.3$. The latter requirement is made without the loss of generality as long as the maximum redshift of the EM survey comfortably exceeds the maximum redshift of the GW survey. Note that our assumption that the redshift distribution of GW events matches the EM data out to $z=0.3$ may not be satisfied in reality, but could be modeled out with a further analysis.

Each selected GW event's redshift is then converted into a true luminosity distance, $\dgw$, by assuming a true Hubble constant of $H_{0{,\rm true}} = 70\kmsMpc$. This value of $H_{0,{\rm true}}$ is what we expect to recover in all cases if the method is not biased. Each GW event distance weights the parameter space of allowed values of $z$ and $H_0$ according to
\begin{equation}
    p(\dgw|\dl(z,H_0))\propto \exp\bigg{[}-\frac{1}{2}\bigg{(}\frac{d_L(z,H_0)-\dgw}{\sigma_{\dl}}\bigg{)}^2\bigg{]}.
    \label{eq:p_dgw}
\end{equation}
This equation would be the likelihood for a single GW event with a single EM candidate whose redshift is perfectly known. When we account for many EM candidates, the likelihood includes weighted galaxy votes from $p(z)$ and normalizes by selection functions, as shown in Eq.~\ref{eq:pH0}

\subsection{Bayesian Analysis}

Applying Bayes' theorem, we can easily evaluate the posterior on $H_0$; it is given by the likelihood in Eq.~(\ref{eq:p_dgw}) convolved with the distribution $p(z)$ as discussed above, and combined with some priors:
\begin{equation*}
p(H_0 | \dgw,\dem) = \frac{\int p(\dgw|\dl(z,H_0))p(z){\rm d}z}{\int P_{\rm det}^{\rm EM}(z) P_{\rm det}^{\rm GW}(\dl(z,H_0)) p(z){\rm d}z} \end{equation*}
\begin{equation}
 \propto \frac{\int p(\dgw|\dl(z,H_0))p(z){\rm d}z}{\int\limits_0^{d_{L,{\rm max}(H_0)}} p(z){\rm d}z}.
\label{eq:pH0}
\end{equation}
where $P_{\rm det}^{\rm EM}(z)$ and $P_{\rm det}^{\rm GW}(\dl(z,H_0))$ are selection terms that account for the detectability of galaxies and GW events. For a full description of these terms we recommend Ref.~\cite{Chen:2017rfc,Fishbach:2018gjp}.

Here $P_{\rm det}^{\rm EM}(z)$ is the probability that a galaxy at redshift $z$ will be captured in the galaxy catalog. Consistent with \cite{Chen:2017rfc,Fishbach:2018gjp} adopt a simple form
\begin{equation}
    P_{\rm det}^{\rm EM}(z) \propto \mathcal{H}(z_{\rm max}-z)
\label{eq:pem}
\end{equation}
where $\mathcal{H}$ is the Heaviside step function and $z_{\rm max} = 0.3$. Similarly, $P_{\rm det}^{\rm GW}(\dl(z,H_0))$ is the probability that a GW event at a distance $d_L$ will be captured in the GW catalog. We likewise simplify this term as
\begin{equation}
    P_{\rm det}^{\rm GW}(\dl(z,H_0)) \propto \mathcal{H}(d_{L,{\rm max}}-d_L)
\label{eq:pgw}
\end{equation}
where our previous choices of $z_{\rm max} = 0.3$ and $H_{0,{\rm true}} = 70\hinvmpc{}$ give us a distance cutoff of approximately $1570\hinvmpc$. This distance remains fixed and corresponds to different redshift cutoffs as a function of $H_0$. Throughout this analysis we adopt the flat priors $z\in[0,2]$ and $H_0\in[40,100]$. We combine GW events by taking the product of each individual event's posterior given in Eq.~(\ref{eq:pH0}).

\begin{figure}[t]
    \includegraphics[width=\linewidth]{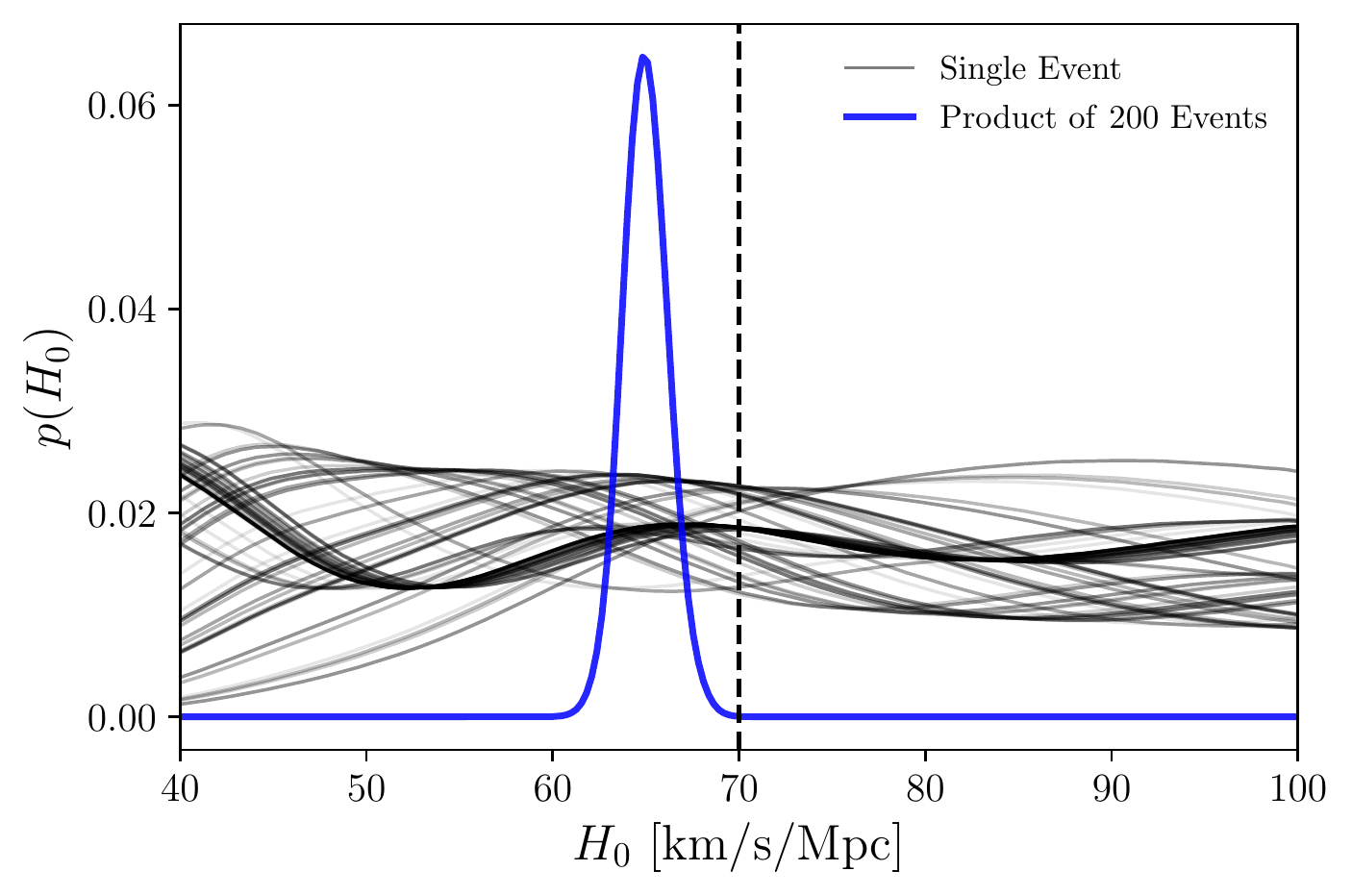}
    \caption{Individual posteriors on $H_0$ from 200 GW events (black) and their product (blue; scaled by 1/5th for clarity). We adopt EM data from the MICECAT catalog that lie in observing direction 1, with an opening angle $\theta = 1^\circ$, and assuming the distance error $\sigma_{d_L} = 10\%$; see text for details. }
    \label{fig:fiducial}
\end{figure}

\section{Results}\label{sec:results}

Fig.~\ref{fig:fiducial} shows the posterior on $H_0$ from 200 GW events, each with $\sigma_{d_L} = 10\%$. The coloring is consistent, so the blue product curve corresponds to the blue $p(z)$ from Fig.~\ref{fig:pz}. We also show the 200 individual GW event posteriors that contribute to the product in grey. These single-event posteriors are not shown in future plots, but are nonetheless part of the process for each forthcoming result. Clearly, any single event posterior is unlikely to be a good indicator of where the result will converge with many more events. More disconcertingly, we also find that even combining 200 events does not necessarily recover $H_{0,{\rm true}}$, even with $1^\circ$ angular localization and $10\%$ distance uncertainty (see a typical example in Fig.~\ref{fig:fiducial}). To see if this problem persists more generally, we now open up the space of analysis choices and data realizations.

\begin{figure*}[t]
    \includegraphics[width=\linewidth]{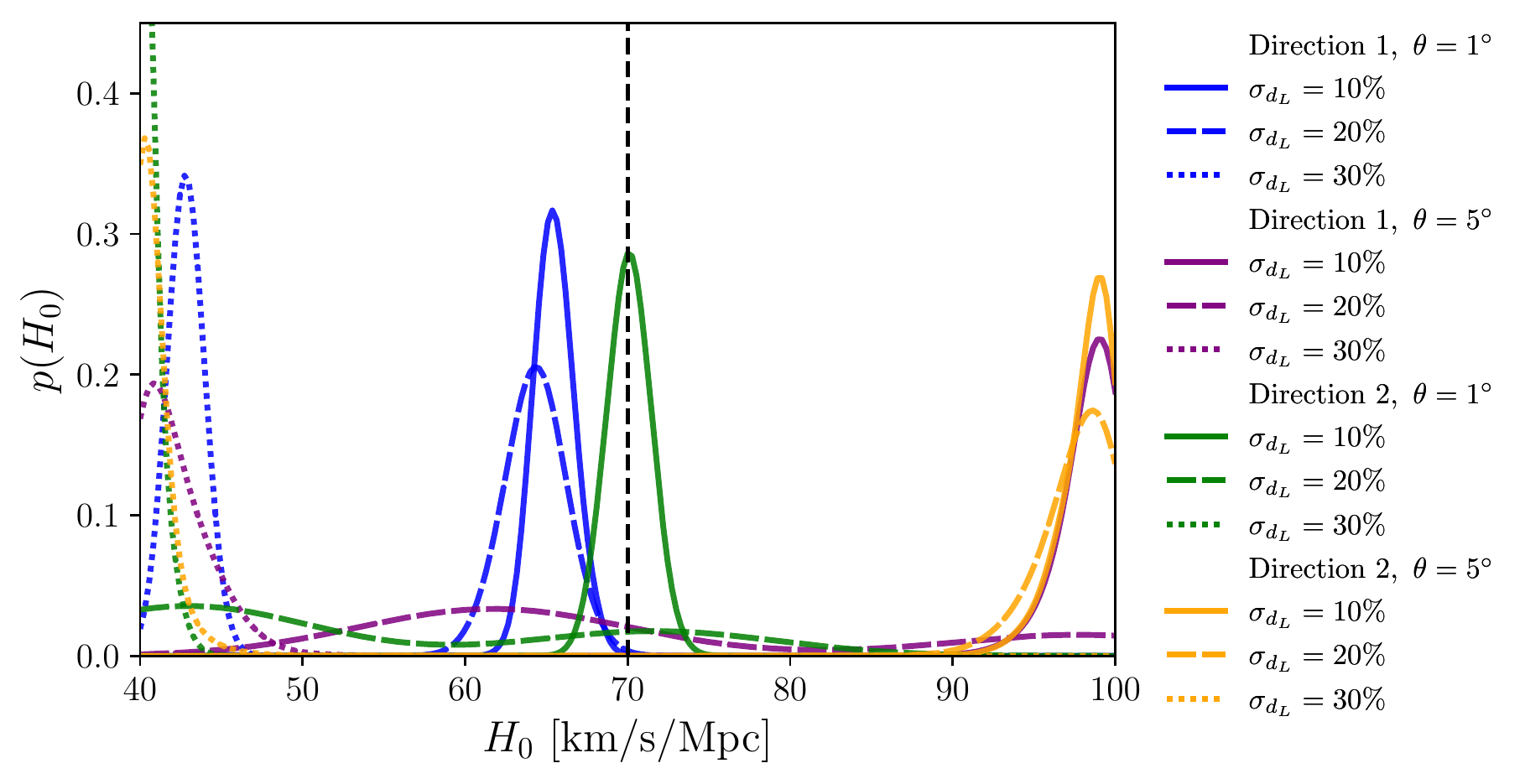}
    \caption{Posteriors on $H_0$, assuming 200 GW events, for different cases of observing direction, opening angle, and distance uncertainty. Curves of the same color share the same set of GW events in the same region of the (simulated) sky, while different line styles represent different distance uncertainties. While it is possible to recover the input value of $H_0$, the peak of the posterior is generally biased depending on distance uncertainty and observing direction. See Sec.~\ref{sec:results} for a further explanation of these results.
    }
    \label{fig:pH0_sigma}
\end{figure*}

Figure \ref{fig:pH0_sigma} shows our principal results. Assuming 200 GW events, we show the posteriors on $H_0$ for various combinations of selecting each of: two MICECAT observing directions, two opening angles (around each direction), and three distance uncertainties. 
The observing-direction and opening-angle combinations are denoted with different colors, while the distance uncertainties are encoded with different line types.

The main finding from Fig.~\ref{fig:pH0_sigma} is that significant biases in the $H_0$ posteriors can be expected.  
While it is \textit{possible} to recover an unbiased value of $H_0$, it is by no means guaranteed, and the bias is generally much larger than the inferred statistical error from the posterior. 

To understand this better, note that the method essentially relies on matching features in the EM galaxy distribution to features in the GW event distribution, and therefore the prominence of these features dramatically affects the results. In Fig.~\ref{fig:pz} there is a noticeable difference in the fluctuations of $p(z)$ between the $\theta = 1^\circ$ cases (larger fluctuations) and the $\theta = 5^\circ$ cases (smaller fluctuations). It is then no surprise that in Fig.~\ref{fig:pH0_sigma} the $\theta = 5^\circ$ cases are less reliable at recovering $H_{0,{\rm true}}$ than the $\theta = 1^\circ$ cases. While these $\theta$ values represent opening angles on the sky (here, in the MICECAT catalog), they serve to make the more general point that smoother $p(z)$'s create less reliable dark-siren measurements of $H_0$.

Furthermore, results differ even for two cases with identical choices of well-localized events --- $\theta = 1^\circ$ and $\sigma_{\dl} = 10\%$, shown in the solid blue and green curves of Fig.~\ref{fig:pH0_sigma}. These cases only differ in observing direction, and yet the differences in $p(z)$ between the two regions of the sky are enough to shift the peak of $p(H_0)$.

Lastly, the results are highly dependent on the distance uncertainty, which  appears in the numerator of Eq.~(\ref{eq:pH0}) as a means of telling us how harshly to punish combinations of $z$ and $H_0$ that do not combine to match the observed GW event distance $\dgw$. Increasing $\sigma_{\dl}$ has a similar qualitative effect as smoothing $p(z)$ by giving more weight to galaxies farther from the measured $\dgw$, which decreases the impact of clusters near $\dgw$ that the method relies on. However, it causes the posterior to peak at the opposite end of the $H_0$ prior because the smoothing is not compensated for in denominator like it is when $p(z)$ itself is changed.
\begin{figure}[t]
    \includegraphics[width=\linewidth]{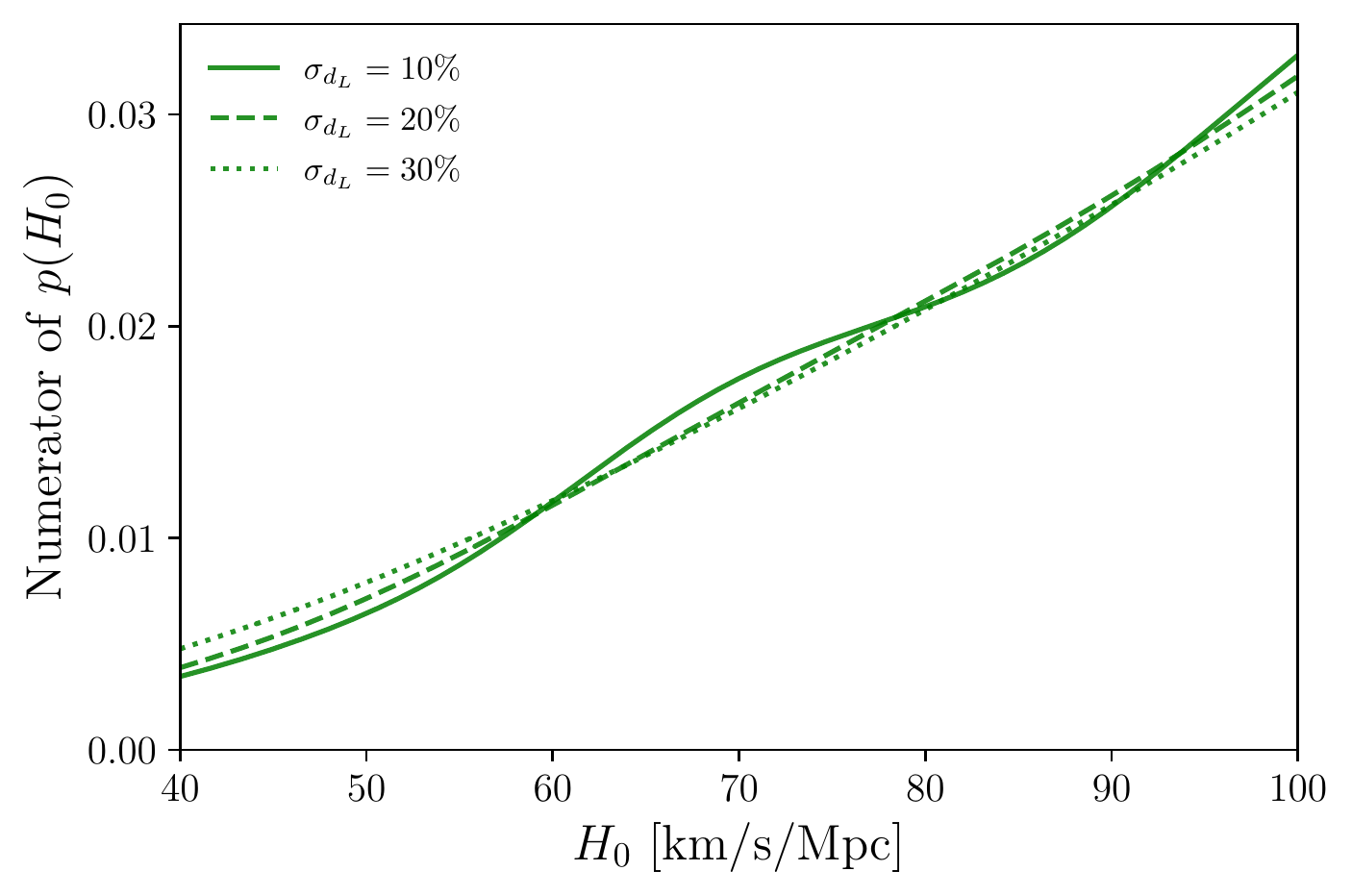}
    \caption{The effect of distance uncertainty on the numerator of Eq.~\ref{eq:pH0}. These curves are the product of 200 GW events and they are numerators of the posteriors that are the green curves in Fig.~\ref{fig:pH0_sigma}. The features are washed out as the distance uncertainty increases.}
    \label{fig:numerators}
\end{figure}

\section{Discussion and Conclusions}\label{sec:concl}

We have studied the statistical method based on dark sirens --- gravitational-wave (GW) events with distance measurements but no electromagnetic (EM) counterpart identification --- to constrain the Hubble constant. To do so, we set up a simulated analysis based on the EM (galaxy) sample adopted from the MICECAT light-cone numerical simulation. We assumed that the GW events follow the underlying EM distribution, and made other specifications that roughly follow expectations from near-future GW data. Throughout, we did not assume or  introduce any systematic errors in the data or the analysis, so our results can be considered a best-case scenario.

As is perhaps well-known but nevertheless worth emphasizing, without any features in the distribution of galaxies $p(z)$, it is manifestly impossible to constrain the Hubble constant using dark sirens. The reason is the irreducible degeneracy between the GW sources' redshifts and the Hubble constant. Assuming for the moment the $z\ll 1$ limit for simplicity, we have
\begin{equation}
    d_L(z, H_0) \simeq \frac{cz}{H_0}.
\label{eq:dl}
\end{equation}
The perfect degeneracy between $z$ and $H_0$ implies that, in the featureless $p(z)$ scenario, even a perfect measurement of $d_L$ cannot lead to a constraint on the Hubble constant because the individual galaxy redshifts are not available for dark sirens. 

It has been argued in the literature that a realistic galaxy distribution would effectively break this degeneracy. Intuitively, the ``bumps" in the distribution of galaxies $p(z)$ --- sampled by GW events that we optimistically assume come from the same distribution --- would effectively serve as standard rulers. Namely, each such bump would effectively select the redshift of that feature, and hence break the degeneracy in Eq.~(\ref{eq:dl}) and constrain the Hubble constant. 
Quantitatively, the individual events' posteriors on $H_0$ will be very broad, but are expected to combine to give the total posterior that is much narrower (see Fig.~\ref{fig:fiducial}).

Unfortunately, we find that this expectation is not borne out by the analysis. The individual events' posteriors do combine into a much more precise constraint, but one that is generally highly biased relative to $H_{0,{\rm true}}$ (see Fig.~\ref{fig:pH0_sigma}). Worryingly, even for identical statistical properties of the EM dataset and GW events, the $H_0$ posteriors vary dramatically depending on the realization of the EM sample (corresponding to direction 1 or 2 in Fig.~\ref{fig:pH0_sigma}). The results also unsurprisingly get more biased when the GW distance error increases, but we find they are generally biased even for an optimistic case of 10\%  error per event.
Overall, these results show that constraints from a large sample of GW dark-siren events are biased in a way that depends on the GW sample characteristics as well as the realization of the underlying EM sample.

What about constraints from a much smaller number of GW events --- say, a single event (as in \cite{Soares-Santos:2019irc})? It has been hoped that such a constraint, while statistically weak, is nevertheless reliable and bound to get stronger with more events. However, a simple inspection of Fig.~\ref{fig:fiducial} shows that this is not the case. The single event curves peak over a wide range of $H_0$ values, and while the product of 200 curves does produce a more statistically significant result, it is not guaranteed to recover $H_{0,{\rm true}}$.

Our numerical experiments have shown that the $H_0$ bias slowly goes away in the limit of large clustering. Specifically, placing most of the galaxies in a narrow redshift range (hence creating a large and unrealistic ``bump" in $p(z)$) does slowly move the posteriors toward $H_{0,{\rm true}}$. This trend  is intuitively best understood in the extreme thought example assuming that all GW events were in a single cluster; a knowledge of the redshift of this cluster would then effectively make redshift of all events known, essentially turning them into \textit{bright} sirens. In a toy example with a rather large amount of clustering (not shown in any of our plots), we do confirm that the $H_0$ posteriors become unbiased.
Similarly, reducing the distance errors to near zero also reduces the $H_0$ bias. 

Another concern is the one we have \textit{not} studied in this paper, which is a mismatch between the galaxy dataset and the underlying population of GW sources. The physical phenomena that govern GW events take place on scales some ten orders of magnitude smaller than those well understood by first-principles theory ($\sim$Mpc). For example, accretion and merger history probably play a role in determining the likelihood that a halo of a given mass is a host of GW sources, yet such properties are notoriously challenging to constrain observationally.

In conclusion, we find that the statistical GW method has significant challenges to overcome in order to become a reliable probe of the Hubble constant. 

\section{Acknowledgments}
We sincerely thank Jonathan Gair for detailed useful feedback on an earlier version of this manuscript, and a lengthy exchange about how to implement the analysis. We would like to thank Jim Annis, Maya Fishbach, Daniel Holz, Antonella Palmese, Keith Riles, Marcelle Soares-Santos, Rachel Gray, Simone Mastrogiovanni, Suvodip Mukherjee, Archisman Ghosh, and Nicola Tamanini for discussions and critical feedback. Our work has been supported in part by NASA under contract 19-ATP19-0058. DH has additionally been supported by DOE under Contract No.\ DE-FG02-95ER40899 an NSF under contract AST-1812961. DH thanks the Humboldt Foundation for support via the Friedrich Wilhelm Bessel award, and the Max-Planck Institute for Astrophysics for hospitality while a large chunk of this work was carried out.

\bibliographystyle{unsrt}
\bibliography{refs}

\end{document}